\documentstyle[12pt,epsfig]{ioplppt}

\def\bxi{\mbox{\boldmath $\xi$}}
\def\bpsi{\mbox{\boldmath $\psi$}}

\def\bneta{\mbox{\boldmath $\eta$}}

\def\bm{{\bf m}}

\begin{document}
\jl{1}
\title{The Ashkin-Teller neural network near saturation}
\author{D Boll\'{e}\dag\ftnote{3}{Also at Interdisciplinair 
Centrum voor Neurale Netwerken, K.U.Leuven, Belgium\\e-mail:
desire.bolle@fys.kuleuven.ac.be, piotr.kozlowski@fys.kuleuven.ac.be} and 
P Koz{\l}owski\dag\ddag~\S}
\address{\dag Instituut voor Theoretische Fysica,
K.U. Leuven, B-3001 Leuven, Belgium}
\address{\ddag Computational Physics
Division, Institute of Physics, A. Mickiewicz University,
PL 61-624 Pozna\'{n}, Poland}

\begin{abstract}
The thermodynamic and retrieval properties of the Ashkin-Teller neural 
network model storing an infinite number of patterns are examined 
in the replica-symmetric mean-field approximation. In particular, for 
linked patterns temperature-capacity phase diagrams are derived for
different values of the two-neuron and four-neuron coupling strengths.   
This model can be considered as a particular non-trivial generalisation 
of the Hopfield model and exhibits a number of interesting new features. 
Some aspects of replica-symmetry breaking are discussed.
 \end{abstract}
\maketitle

\section{Introduction}
The Ashkin-Teller neural network (ATNN) studied recently by us in the case 
of loading of a finite number of patterns (\cite{atnl} and references
therein) can be considered as a non-trivial generalisation of the Hopfield 
model \cite{hop,amit} to allow for different types of neurons. 
It can be seen as a model consisting of two Hopfield models 
interacting through a four-neuron term, or it can be interpreted as a
neural network with two types of neurons at each site having different
functions.

Some of the underlying neurobiological motivation for the introduction
of different types of neurons is the fact that there exist areas in the
brain which react to two different kinds of dependent stimuli in such a
way that the response to particular combinations of these stimuli is
stronger than the response to others \cite{KI}. Furthermore, in
neuropsychological studies on amnesia it has become appreciated that
memory is composed of multiple separate systems which can store
different types of information, e.g., information based on skills and
information based on specific facts or data \cite{S}. 
In addition, the classical Ashkin-Teller model on which our ATNN model
is based, appears in studies of disordered systems when the 
disorder evolves on a time scale that can be tuned \cite{C}.
Finally, two different types of neurons show up naturally when
considering the Hopfield model with synchronous dynamics
\cite{amit,FK,VH}. 

Storage and retrieval of embedded patterns built from these two types of
neurons with different degrees of (in)dependence have been discussed in 
\cite{atnl} for finite loading, i.e. the capacity $\alpha$, defined as the
number of embedded patterns, $p$, versus the size of the system, $N$, is
taken to be zero. In that study we wanted to find out, e.g.,  whether the
four-neuron interaction between the two types of neuron can improve the
retrieval process. And indeed, it has been seen that interesting
retrieval behaviour emerges, especially for linked patterns, e.g., different
types of Mattis states occur, a higher retrieval transition temperature
is possible, the quality of retrieval is improved through a bigger
overlap. Therefore, it is interesting to investigate the
properties of this model for infinite loading of patterns, i.e., for
$\alpha \neq 0$. 
This is the purpose of the present paper.

After describing the
ATNN-model in Section~2 we use the standard replica mean-field theory
approach in order to write down its free energy in Section~3. Assuming 
replica 
symmetry we then obtain the fixed-point equations for the relevant order
parameters. Thermodynamic and retrieval properties are discussed in
Section~4. In particular, Section~4.1 investigates the behaviour of the
model at zero temperature for equal two- and four-neuron coupling 
strengths in the Hamiltonian . In this case also the entropy is calculated
and found to be negative. This leads to a discussion about the size of
the replica-symmetry breaking and re-entrance behaviour. Section~4.2
presents phase diagrams in the temperature-capacity plane for different 
values of the two- and four-neuron coupling strengths. 
A number of new features in these diagrams show up. Also the maximal
information content of the network is obtained as a function of these
couplings. Where relevant, the
results are compared with the Hopfield model \cite{hop,amit} and with the 
four-state Potts model \cite{kanter,potts,jacky}. Finally, Section~5
presents some concluding remarks.

\section{The model}
We consider a network of $N$ sites each with two different types of binary 
neurons $s_i$ and $\sigma_i, i=1, \ldots, N$. The interactions between the 
neurons are determined by the infinite-range Hamiltonian
\begin{equation}
   H=-\frac{1}{2} \sum_{i \neq j} \left[J_{ij}^{(1)} s_i s_j +
     J_{ij}^{(2)} \sigma_i \sigma_j + J_{ij}^{(3)} s_i s_j \sigma_i
     \sigma_j\right]       \, .  
           \label{ham}
\end{equation}

In this network storage and retrieval of two different types of patterns, 
$\bxi_i=\{\xi_i^\mu\}$ and $\bneta_i=\{\eta_i^\mu\}, \mu=1,\ldots,p$ 
are enabled by a Hebb-like learning rule for the couplings 
\begin{equation}
\fl \hspace*{0.5cm}  J_{ij}^{(1)}= \frac{1}{N} J_1 
           \sum_{\mu=1}^p \xi_i^\mu \xi_j^\mu, \quad
     J_{ij}^{(2)}= \frac{1}{N} J_2 \sum_{\mu=1}^p \eta_i^\mu \eta_j^\mu,
     \quad
     J_{ij}^{(3)}= \frac{1}{N} J_3 \sum_{\mu=1}^p \xi_i^\mu
     \eta_i^\mu\xi_j^\mu\eta_j^\mu~.
        \label{hebb}
\end{equation}

The patterns $\bxi_i$ and $\bneta_i$ are supposed to be independent 
identically distributed random variables (i.i.d.r.v.) taking the values 
$+1$ or $-1$ with equal probability. 
All $J_\nu, \nu=1,2,3$ are non-negative.

First, we remark that the above definition of $J_{ij}^{(3)}$ is equivalent
to the case of linked patterns considered in \cite{atnl}. Second, we repeat 
that this model can be considered as an assembly of two Hopfield models
(when $J_3=0$) interconnected via a four-neuron interaction (when $J_3\neq
0$).

\section{Replica-symmetric mean-field theory}
In order to determine the thermodynamic and retrieval properties of the 
model we calculate the quenched free energy per site within the
replica-symmetric mean-field theory approach. Using standard techniques
\cite{amit} we arrive at  
\begin{eqnarray}
\fl \beta f(\beta)=\frac{1}{2}\sum_{\nu=1}^3\left\{\sum_{\mu=1}^c K_{\nu} 
          (m_{\nu}^{\mu})^2+
          \frac{3}{2}\alpha \left[ K_{\nu}+K_{\nu}^2 r_{\nu} (1-q_{\nu})-
       \frac{K_{\nu}q_{\nu}}{1-K_{\nu}(1-q_{\nu})} \right.\right.
         \nonumber\\
 \fl \hspace*{1cm} 
    + \left. \left.\ln [1-K_{\nu}(1-q_{\nu})]
    \rule{0.cm}{0.6cm}\right]\rule{0.cm}{0.7cm}\right\}-
                \int\prod_{\gamma=1}^3 {\rm D}z_\gamma
              \left\langle\!\left\langle 
	 \ln \left[ 4\prod_{\nu=1}^3 \cosh a_{\nu}(1+
	         \prod_{\nu=1}^3\tanh a_{\nu})
             \right]\right\rangle\!\right\rangle ~~,
	 \label{free}
\end{eqnarray}
with $\beta=1/T$ the inverse temperature measuring the noise level in the
system, $\alpha$ the capacity defined as the number of patterns per 
number of couplings per spin, i.e., $\alpha =2p/3N$ and where
\begin{eqnarray}
   a_{\nu}&=& K_{\nu}\left(\sum_{\mu=1}^c m_{\nu}^{\mu}\psi_\nu^\mu +
     z_\nu\sqrt{\frac{3}{2}\alpha r_\nu}\right)~,
       \label{defk2} \\
   K_\nu &=& \beta J_\nu~,~~\psi_1^\mu=\xi^\mu~,~~\psi_2^\mu=\eta^\mu~,
         ~~\psi_3^\mu=\xi^\mu\eta^\mu~. 
	 \label{defk1}   
\end{eqnarray}
Here the index $\mu=1,...,c$, $c$ finite, labels the condensed patterns, 
the index $\nu =1,2,3$, $\langle\!\langle \cdots \rangle\!\rangle$
indicates the  
average over the embedded patterns and D$z_\gamma$ denotes the Gaussian 
measure D$z_\gamma = {\rm d}z_\gamma {(2\pi)}^{-1/2}
exp({-z_{\gamma}^2/2})$.
In (\ref{free})-(\ref{defk2}) the $\bm_{\nu} = \{ m_\nu^\mu \}$ are the 
overlap order parameters between the pattern $\bpsi_\nu$ and the network
state $\{S_i^\nu \}$, the $q_\nu$ represent the Edwards-Anderson (EA) order
parameters
with their conjugate variables $r_\nu$ (the mean-square random overlap
with the non-condensed patterns), viz.
\begin{eqnarray}
 \fl \hspace*{0.5cm}  \bm_{\nu}=\left\langle\!\left\langle
       \frac{1}{N}\sum_{i=1}^N\left\langle S_i^\nu\right\rangle\bpsi_i
                \right\rangle\!\right\rangle ~,
    ~~q_\nu =\left\langle\!\left\langle \frac{1}{N}\sum_{i=1}^N
          \left( \left\langle S_i^\nu\right\rangle\right)^2 
	     \right\rangle\!\right\rangle~,
    ~~r_\nu =\frac{2}{3 \alpha}\sum_{\mu =c+1}^p\left\langle\!\left\langle 
          \left( m_\nu^\mu\right)^2\right\rangle\!\right\rangle
\end{eqnarray}
with $S_i^1=s_i, S_i^2=\sigma_i, S_i^3= s_i \sigma_i$ and where  
$\langle \cdots \rangle$ denotes the thermal average. 

The phase structure of the network is determined by that solution of the
set of fixed-point equations
\begin{eqnarray}
   \bm_\nu =\int\prod_{\gamma =1}^3 {\rm D}z_\gamma
         \left\langle\!\left\langle
        \bpsi_\nu\frac{\tanh a_\nu +\tanh a_\delta \tanh a_\rho}
        {1+\tanh a_\nu \tanh a_\delta \tanh a_\rho}
               \right\rangle\!\right\rangle
       \label{m}\\
   q_\nu=\int\prod_{\gamma =1}^3 {\rm D}z_\gamma
       \left\langle\!\left\langle\left(
           \frac{\tanh a_\nu +\tanh a_\delta \tanh a_\rho}
             {1+\tanh a_\nu \tanh a_\delta \tanh a_\rho}\right)^2
                \right\rangle\!\right\rangle
       \label{q}\\
   r_\nu =q_\nu\left[ 1-K_\nu (1-q_\nu)\right]^{-2}  
       \label{r}
\end{eqnarray}
which maximizes $-\beta f(\beta)$. Here $\nu, \delta, \rho =1,2,3$ and
our convention is that they have to be taken different in the equations
above.

Solving this set of equations is a tedious task. In the following sections
we discuss these solutions that are important for the
thermodynamic and retrieval properties of the model, i.e., Mattis
retrieval states and spin-glass states. At this point we remark that 
in the $\alpha =0$ limit the replica-symmetric solution of the ATNN model
presented above is completely consistent with the corresponding result 
obtained in \cite{atnl}.

\section{Results}
The solution of the fixed-point equations for the ATNN neural network
model presented in the previous section depends, of course, on the 
values of the coupling strengths $K_\nu$. Without loss of generality
we put $K_1=K_2$ in the sequel. Such a choice is very often used in 
the literature concerning classical (\cite{kam} and references 
therein) and  spin-glass \cite{sher} Ashkin-Teller models.
This model contains as particular limits the Hopfield network for any 
two $K_\nu$ equal to zero,  the four-state clock network \cite{cook}
(or, equivalently, two independent Hopfield networks) for any $K_\nu=0$
and a model strongly resembling the four-state Potts network
\cite{kanter,potts} for $K_1=K_2=K_3$. Remark, however, that only the
strength of the couplings is the same but, in general, $J_{ij}^{(1)}
\neq J_{ij}^{(2)} \neq J_{ij}^{(3)}$  (compare \cite{sher}).
Therefore, some details of the phase diagram compared with
those of the Potts network may be different.  

The type of solutions we are interested in, at first instance, are
Mattis solutions, for which not more than one component of each order
parameter $\bm_\nu$ is non-zero. 
As in \cite{atnl} we thereby distinguish two 
types of Mattis states. Simple Mattis states where only the same 
components of the $\bm_\nu$ are non-zero, e.g., $\bm_1=(m_1,0,...,0)$, 
$\bm_2=(m_2,0,...,0)$, $\bm_3=(m_3,0,...,0)$, and crossed Mattis states
where the same components of the $\bm_\nu$ are never non-zero, e.g., 
$\bm_1=(m_1,0,...,0)$, $\bm_2=(0,m_2,0,...,0)$,
$\bm_3=(0,0,m_3,0,...,0)$. These states are denoted by $m_1m_2m_3$ in
the sequel. All these Mattis retrieval solutions also have 
non-zero values of the EA order parameters $q_\nu$.
We remark that we have also found Mattis states where only one $\bm_\nu$ 
and the corresponding $q_\nu$ are non-zero. These are, in
fact, Hopfield-like solutions.
Furthermore, we are interested in spin-glass (SG) states characterised by
$\bm_\nu =0$, $q_\nu\neq 0$ (here also Hopfield-like SG states
are possible) and a paramagnetic state with all order parameters 
$\bm_\nu =0$, $q_\nu = 0$.

First it is worthwhile to discuss the most simple case, i.e., zero 
temperature  with equal coupling strengths. Second we present results for
non-zero temperatures and arbitrary couplings.

\subsection{Retrieval properties at zero temperature}
Restricting ourselves first to $K_1=K_2=K_3$ we find that only 
solutions with equal order parameters, $\bm_1=\bm_2=\bm_3$ and
$q_1=q_2=q_3$ exist. Further, considering simple Mattis states at zero  
temperature the set of fixed-point equations (\ref{m})-(\ref{r}) can 
be reduced to a single equation, as in \cite{amit} 
\begin{eqnarray}
  \fl x\sqrt{3\alpha}=
       \int_{-\infty}^\infty {\rm D}z ~{\rm erf} 
        \left(\frac{z}{\sqrt{2}}+2x\right)
   -\int_{-x\sqrt{2}}^\infty {\rm D}z~{\rm erf}
       \left(\frac{z}{\sqrt{2}}+2x\right)\left[1-zx\sqrt{2}\right]
             \nonumber\\
  \fl \hspace*{0.5cm}+ \frac{1}{2}\int_{-\infty}^{-x\sqrt{2}}{\rm D}z  
    \left[ {\rm erf}^2\left(\frac{z}{\sqrt{2}}+2x\right) +
         {\rm erf}^2\left(\frac{z}{\sqrt{2}}\right)\right]\left[
               1-zx\sqrt{2}\right] 
	     -x\sqrt{\frac{2}{\pi}}\left(1+e^{-2x^2}\right)
	     \label{xeq}
\end{eqnarray}
with the new variable $x=m/\sqrt{3\alpha r}$ being different from zero
only when $m\neq 0$. Thus the range in $x$ allowing the existence of 
non-zero solutions of eq.~(\ref{xeq}) determines the critical capacity,       
$\alpha_{c}(0)$, of the ATNN model with equal coupling strenghts
at $T=0$. 
Numerically solving this equation for different values of 
$\alpha$ we find  $\alpha_{c}(0)=0.1839205$. This is higher than the
critical  capacity of the Hopfield model ($\alpha_c(0)= 0.137905566$)   
\cite{steffan,bh} and agrees (up to numerical precision) with the result
obtained in ref.~\cite{kanter} for the four-state
Potts neural network after an appropriate rescaling with the number of
couplings as discussed in \cite{gerl}. (The relation between the
Hopfield and Potts critical capacities is given in eq.~(19) of
\cite{kanter}.)
We remark that taking the coupling strengths $K_\nu$ unequal always
leads to a smaller critical capacity, as we will see in the next subsection.
 
In analogy with the Ashkin-Teller spin-glass \cite{sher} and with
the Hopfield network \cite{amit} we expect replica symmetry to be broken 
for low temperatures. Consequently, we expect some re-entrant SG-behaviour
(see, e.g., \cite{reent}) indicating that the true critical
capacity is greater than its replica symmetric value at $T=0$.
In order to get an idea about this breaking we have calculated the
entropy of the replica-symmetric solution at $T=0$, which for simple
Mattis states reads
\begin{equation}
  S(\alpha)=-\frac{9}{4}\alpha\left[\ln (1-C)+\frac{C}{1-C}\right],~~~~~
        ~C=\lim_{\beta\rightarrow\infty}\beta(1-q)~.
\end{equation}
As expected we find that the value of this entropy is negative. In
particular, $S(\alpha_c(0))=-0.007228$ versus $-0.001445$ for the
Hopfield model \cite{amit}, $-0.003995$ for the three-state Potts model
\cite{potts} and $-0.007212$ for the four-state Potts model
\cite{jacky}.  Furthermore, this entropy vanishes exponentialy with
decreasing $\alpha$ and, in fact, for small $\alpha$ we have that 
$S \approx -\exp(-1/\alpha)$. In the SG-phase $S= -0.91$
at $\alpha_c(0) $ versus $-0.07$ for the Hopfield model \cite{amit} and 
$-0.13$ \cite{potts} for the three-state Potts model.
All this suggests that replica-symmetry   
breaking of the retrieval states of the ATNN model is somewhat stronger
than in the Hopfield model but still weak compared with the SG state
of the ATNN.  Further details on this matter are beyond
the scope of the present work.

\subsection{The phase diagram for non-zero temperatures}
The phase diagram of the ATNN model is obtained by
numerically solving the set of fixed-point equations 
(\ref{m})-(\ref{r}). It is a complicated function of $K_1=K_2$, $K_3$, 
and $\alpha$, the behaviour of which we analyse by looking specifically
at $T-\alpha$ sections for different values of the ratio
$w=K_1/K_3$. 
Furthermore, we discuss the maximal information content of
the network in the full space of couplings.

\begin{figure}
\epsfysize=8cm
\epsfxsize=14cm
\centerline{\epsfbox{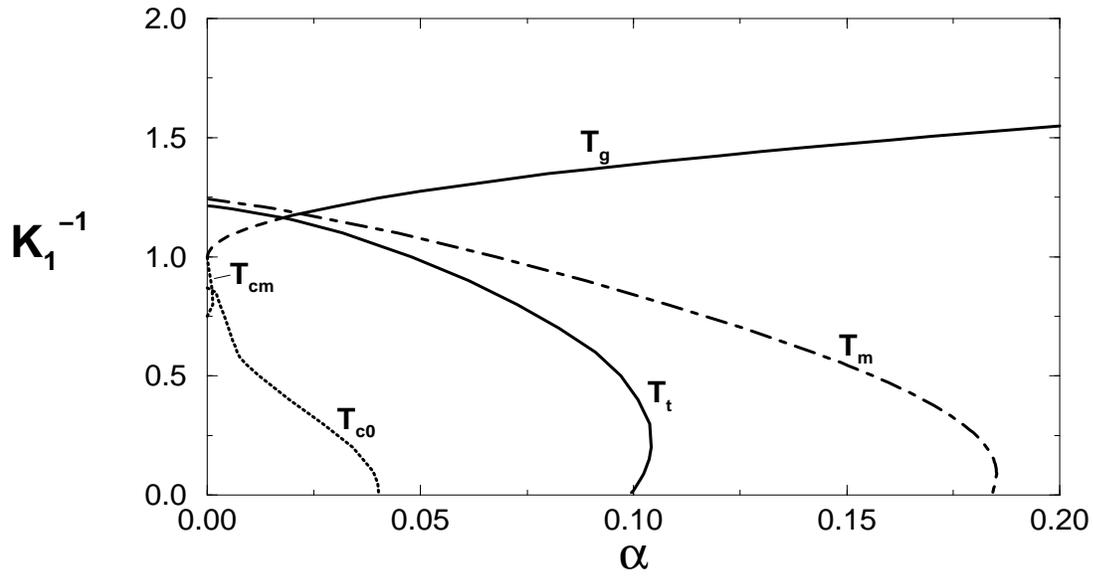}}
\caption{The $T-\alpha$ section of the phase diagram for the ATNN
model with equal coupling strengths ($w=1$). The solid lines
indicate the thermodynamic transitions. The precise meaning of the lines
is described in the text.}
\label{diag}
\end{figure}

An extensive numerical analysis allows us to distinguish essentially two
qualitatively different $T-\alpha$ sections of the full phase diagram.
They are shown in figs.~\ref{diag} and \ref{diag1}.

Several transition lines bordering different phases show up.
The lines $T_m$ and $T_t$ are related to simple Mattis solutions,
the line $T_g$ separates the SG and paramagnetic solutions.
The lines $T_{c0}$ and $T_{cm}$  
are connected with crossed states. Finally, the lines $T_{m1}$, $T_{g1}$
and $T_{t1}$
have to do with the Hopfield-like retrieval and SG solutions
introduced before. We remark that from now on writing that a
solution exists also implies that it is stable within the
replica-symmetric approximation, unless stated otherwise.
  
Let us first discuss fig.~\ref{diag} where the coupling strenghts are all
equal, and hence $w=1$, in more detail. We start looking at high $T$ 
expressed in units of $K_1^{-1}=T/J_1$. The
transition from the disordered paramagnetic phase to the SG-phase 
($\bm_\nu =0$, $q_\nu \neq0$) is continuous and denoted by $T_g$. When
crossing the line $T_m$ simple Mattis retrieval states show up as local
minima of the free energy. At these points the overlap with the embedded
patterns jumps from zero to a finite macroscopic value. So the system
functions as an associative memory and the critical storage capacity for
a given temperature can be read off through that line. 

When further
lowering $T$ the simple Mattis states become global minima of the free
energy. This happens along the line $T_t$ and this thermodynamic
transition is first order. Here we note that the critical lines $T_t$ 
and $T_g$ end in
different temperature points at $\alpha =0$ giving rise to a
``crossover'' region for small $\alpha$ as it occurs in the Potts model 
\cite{potts}. This is related to the fact that
for $\alpha =0$ this ATNN model with $w=1$ has a
discontinuous transition at $T_t$ as shown in \cite{atnl} (see in
particular  fig.~5). In this crossover region the simple Mattis states
(global minima) and the paramagnetic state (local minimum) coexist. 

Finally, at still lower $T$ (and appropriate values of $\alpha$) crossed
states of the type $mmm$ and $mm0$ exist left from the line $T_{cm}$ and
$T_{c0}$ respectively. Their free energy is  bigger than that of the
simple Mattis states. 

We end the discussion of this phase diagram by remarking  that we find
weak re-entrant behaviour both at the lines $T_t$ and $T_m$. The
re-entrance at the line $T_t$ is somewhat stronger and may suggest
\cite{CN}  a greater
replica-symmetry breaking of the SG-solution, in agreement with the
values of the zero-temperature entropy calculated in section~4.1.
Furthermore the greatest capacity of the ATNN model with equal
coupling strengths is obtained for temperature $T=0.09$ and reads 
$\alpha_c(T=0.09)= 0.1851$. 

\begin{figure}
\epsfysize=6cm
\epsfxsize=10cm
\centerline{\epsfbox{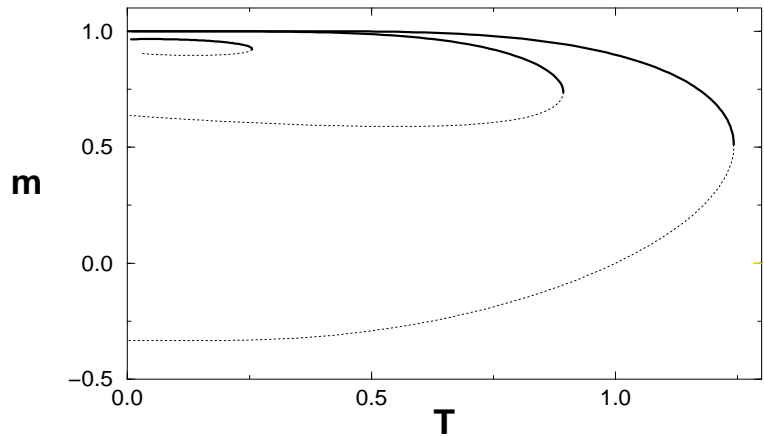}}
\caption{Typical overlap profiles as a function of the temperature
for simple Mattis states of the model with $w=1$ for different
capacities. From bottom to top $\alpha =0,0.09,0.18$. The dotted parts of
the curves are unstable within replica symmetry.}
\label{msimple}
\end{figure}

\begin{figure}
\epsfysize=6cm
\epsfxsize=10cm
\centerline{\epsfbox{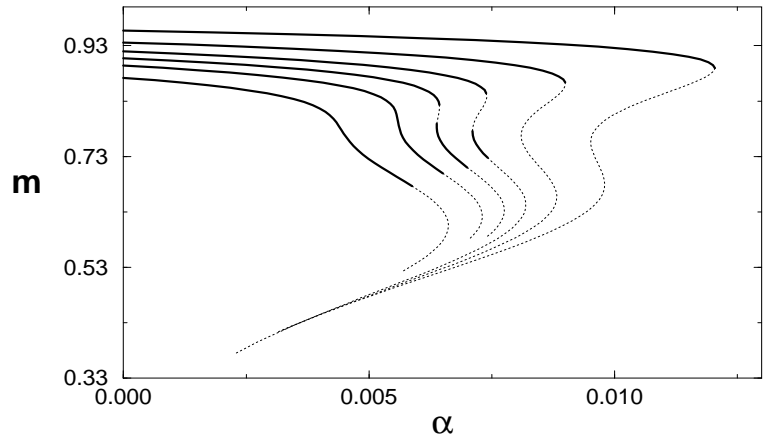}}
\caption{Typical overlap profiles as a function of the capacity for
crossed Mattis states $mm0$ of the model with $w=1$ for different
temperatures. From top to bottom: $T=0.5$, $0.55$, $0.58$, $0.6$,
$0.62$, $0.65$. The dotted parts of the curves are unstable within
replica symmetry.}
\label{mcross}
\end{figure}

\begin{figure}
\epsfysize=8cm
\epsfxsize=14cm
\centerline{\epsfbox{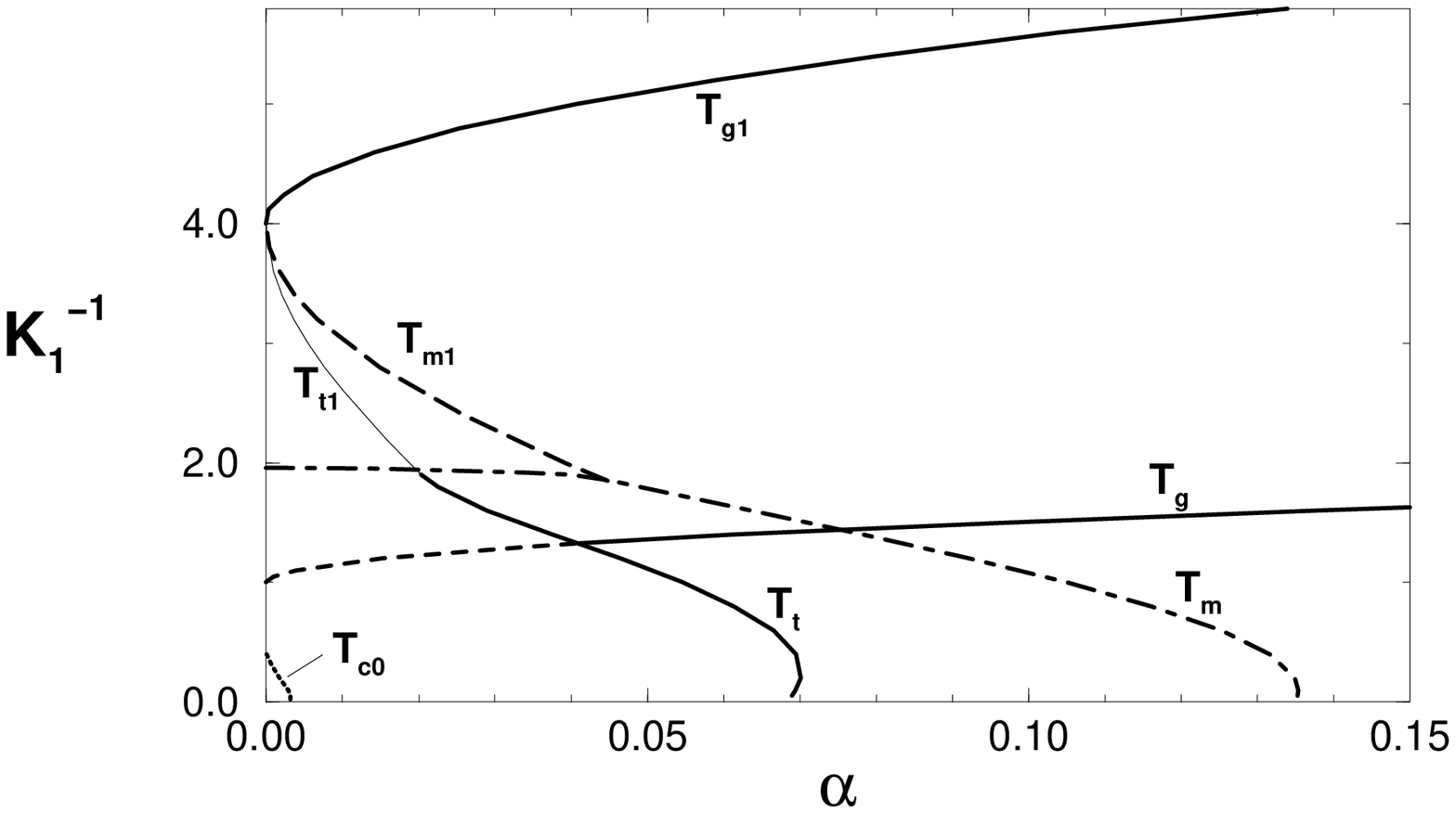}}
\caption{The $T-\alpha$ section of the phase diagram for the ATNN
model with unequal coupling strengths ($w=1/4$). The solid lines
indicate the thermodynamic transitions. The precise meaning of the lines
is described in the text} 
\label{diag1}

\end{figure}

In figs.~\ref{msimple} and \ref{mcross} we present some typical overlap
profiles for simple and crossed Mattis states in this
model indicating the stability of the solutions within the
replica-symmetric approximation (i.e., indicating
where the solution is a minimum of the replica-symmetric free energy).
 We see that, for the same values of $T$
and $\alpha$ the simple states always have the largest overlap. 

Next, we turn to a discussion of the phase diagram for the ATNN 
model with unequal coupling strengths. We explicitly consider the
situation of fig.~\ref{diag1} where $w=1/4$ (meaning that the four-neuron 
term has four times the strength of the two-neuron one).
Starting from high $T$ we first meet the (new) line $T_{g1}$ marking the
continuous transition from the disordered paramagnetic phase to a
SG1-phase where only one of the $q_\nu$ is non-zero, i.e. the
Hopfield-like SG-state, $\bm_\nu =0$, $ q_1=q_2=0, q_3 \neq0$. When
lowering $T$ and taking $\alpha$ sufficiently small the (new) line
$T_{m1}$ appears below which solutions with only one $\bm_\nu$ and      
corresponding $q_\nu$ different from zero, i.e., Hopfield-like Mattis
states  exist as local minima of the free
energy. Below the thermodynamic first order transition line $T_{t1}$
they become global minima of the free energy. These Hopfield phases
exist in the interval $w=[0,0.69]$. 
The rest of the lines which are present are similar to the ones of 
fig.~\ref{diag} whereby we remark that the simple Mattis states are now
of the form $mml$. 
However, crossing the line $T_m$ from  within the Hopfield-like
retrieval phase we observe that the overlap $m_1=m_2$ continuously
increases from zero. Outside this region $m_1=m_2$ immediately jumps
from zero to a finite  macroscopic value as is the case in
diagram~\ref{diag}. 
Furthermore, crossed Mattis states of the type $mml$ have become
unstable for this value of $w$ whereas the crossed states of type $mm0$
exist for all $w > 0$. 

Sections of the full phase diagram for different values of the ratio $w$
are qualitatively not much different and can be explained by taking the 
diagram fig.~\ref{diag} with $w=1$ as a reference.

When $w$ is increasing (starting at the value $1$), meaning that the    
coupling strength of the four-neuron interaction becomes smaller, the
crossover region shrinks since the endpoints of the lines $T_g$ and
$T_m$ come closer. It finally shrinks to zero for $w=\infty$ meaning
that one is left with two independent Hopfield models. The region where
the crossed solutions $mm0$ are found gets larger such that the line
$T_{c0}$ moves to the right until it finally
coincides with the line $T_m$.  The shapes of the transition lines are  
deforming continuously in such a way that they resemble more and more
the Hopfield-model transition lines and finally they become identical
with the latter.

When decreasing $w$  from the value $1$ the phase diagram shown in
fig.~\ref{diag} evolves to the one presented in fig.~\ref{diag1}. The
gap between the lines $T_m$ and $T_g$ for $\alpha =0$ is increasing and
Hopfield-like phases appear. When further decreasing $w$ the gap still
grows until it reaches the maximal value $1$ in units of $K_1^{-1}$ and
the Hopfield  phase occupies a larger part of the phase diagram to become   
completely dominating in the limit $w=0$ where the pure Hopfield model is
found. This means that the lines $T_g$, $T_m$  and $T_t$ shrink and 
disappear as $w \rightarrow 0$ and as a result the lines $T_{m1}$ and
$T_{t1}$ become equal to the corresponding lines of the pure Hopfield
model.

Finally, in order to have an idea about the amount of information stored
in the network we have to take into account both the value of the
capacity and the retrieval overlap. We follow the approach of
\cite{amit2} based upon the Shannon 
entropy \cite{shan}. This leads straightforwardly to the following
information content per number of couplings per spin 
\begin{equation}
    I(\alpha, m_1)=\alpha \left[ \frac12 (1+m_1)\ln (1+m_1)
              + \frac12 (1-m_1)\ln (1-m_1)  \right] \, .
\end{equation} 
We remark that only $m_1$ appears in this formula since for $K_1=K_2$ we
recall that $m_1=m_2$.
We have calculated $I(\alpha, m_1)$ for all $\alpha < \alpha_c$ and
fig.~5 shows its maximal value, $I$,  as a function of  
$K_1^{-1}$ and  $K_3^{-1}$. A few remarks are in order. First, we
find that the maximal information content is reached for values of
$\alpha$ slightly smaller than $\alpha_c$. Second, the greatest
information, i.e. $I=0.1576$, is obtained for equal coupling strenghts.
Third, for $w=\infty$ we find a constant information content independent
of  $J_3$  which is
equal to the one in the Hopfield model at zero temperature.
This Hopfield value is
$0.1213$ (see \cite{amit2}), i.e., $0.0809$ in fig.~5 in view of the
rescaling  with $2/3$ due to the different number of couplings per spin.
Finally, for $w=0$ and $0 \leq K_1^{-1} \leq 2$ the critical capacity is
equal to 
the Hopfield value and zero for $K_1^{-1} >2$ but the information
content is continuously going to zero as $K_1^{-1}$ tends to $2$. This
is due to the fact that the overlap $m_1=m_2$ corresponding to maximal  
information goes continuously to zero.

\section{Concluding Remarks}
In this paper we have studied the thermodynamic and retrieval properties
of the ATNN neural network model storing an infinite number of patterns
using replica-symmetric mean-field theory. This analysis extends our
results for this model in the case of low loading \cite{atnl}.

We have derived the free energy and fixed-point equations for the
relevant order parameters and have obtained the phase diagram for
arbitrary values of the coupling strength of the two- and
four-neuron interactions. This model can be considered as the sum of two 
interacting binary networks such that it contains the Hopfield model
when some of the couplings go to zero. For equal coupling
strengths this model strongly resembles the four-state Potts neural
network.  
   
The storage capacity defined as the number of patterns per number of
couplings per spin is maximal for equal coupling strenghts and equal
to $0.1851$. In that case the maximal information content per
coupling is $0.1576$ versus $0.1213$ for the Hopfield model. 

Different retrieval characteristics can be distinguished in the phase
diagram depending on the parameters of the network. First, the 
two types of embedded patterns are retrieved and the retrieval
quality is enhanced by the interaction between the networks. These
retrieval states are the simple Mattis states. Second, the 
two types of embedded patterns are retrieved but with lower
precision because of the presence of this interaction. These retrieval
states are the crossed states of the type $mm0$ and $mmm$. They do not
play an important role in the thermodynamics of the model.
Third, a pure Hopfield-like phase is present indicating the retrieval of 
only one kind of pattern.

These results allow us to mention some possible biological
applications of this model. Patterns related with two different order
parameters may be seen as a representation for two different stimuli,
e.g., two visual ones as shape and colour, or two different ones as
colour and flavour. When these two stimuli are somehow linked, e.g.,
through the fact that they were experienced at the same time, it is very
likely that they will be recalled together, what corresponds to a
suggestion in ref.~\cite{KI} that particular
combinations of stimuli are favoured.

Finally, we remark that there is some re-entrant behaviour in the phase
diagram at low temperatures which points at replica-symmetry breaking
effects. Calculation of the zero-temperature entropy supports the
expectation that this breaking is weak.

In brief, the presence of a four-neuron coupling term in the ATNN
model not only enhances the quality of pattern retrieval (by giving
a larger overlap at higher temperatures), as present
already in the case of low loading \cite{atnl}, but it also increases the 
critical capacity and the maximal information content in comparision
with the Hopfield model.

\section*{Acknowledgements}
This work has been supported in part by the Research Fund of the K U
Leuven (grant  OT/94/9). Both authors are indebted to  Marc Van
Hulle of the Neurophysiology Department of the K.U.Leuven for
interesting discussions concerning the possible biological relevance of
this model and to  J.~Huyghebaert for useful comments on the
replica-symmetry breaking aspects of this work. They would like to thank
the Fund for
Scientific Research-Flanders (Belgium) for financial support.


\begin{figure}
\epsfysize=5cm
\epsfxsize=9cm
\centerline{\epsfbox{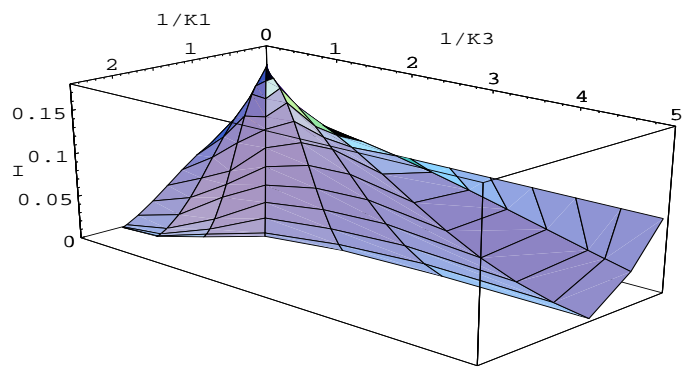}}
\caption{The maximal information content of the ATNN model as a function
of $K_1^{-1}$ and $K_3^{-1}$.}
\label{diag3d}
\end{figure}

\end{document}